\documentclass[12pt]{article}

\usepackage{amsmath,amsgen,amstext,amsbsy,amsopn,amsthm,amssymb,epsf,bbm,
graphics,epic,color,rotating}

\usepackage{epsfig}

\usepackage{graphics,graphicx,wrapfig}

\newcommand{\tmop}[1]{\ensuremath{\operatorname{#1}}}
\newcommand{\tmmathbf}[1]{\ensuremath{\boldsymbol{#1}}}
\newtheorem{theorem}{Theorem}
\newcommand{\tmem}[1]{{\em #1\/}}

\begin{document}

\bibliographystyle{unsrt}

\title{On the Kert\'esz line: \\
Thermodynamic versus Geometric Criticality}

\author{Ph. Blanchard$^1$, D. Gandolfo$^2$, L. Laanait$^3$, \\
J. Ruiz$^2$, and H. Satz$^1$}


\date{}
\maketitle

\begin{abstract}
The critical behaviour of the Ising model in the absence of an external 
magnetic field can be specified either through spontaneous symmetry breaking 
(thermal criticality) or through cluster percolation (geometric criticality). 
We extend this to finite external fields for the case of the Potts' model,
showing that a geometric analysis leads to the same first order/second
order structure as found in thermodynamic studies. We calculate the
Kert\'esz line, separating percolating and non-percolating regimes,  
both analytically and numerically for the Potts model in presence of 
an external magnetic field.

\end{abstract}

pacs: 05.50.+q,64.60.C,75.10.H,05.70.Fh,05.10Ln

\footnotetext[1]{%
Fakult\"at f\"ur Physik,
Universit\"at Bielefeld,
D-33501, Bielefeld, Germany}
\footnotetext[2]{%
Centre de Physique Th\'eorique, UMR 6207, 
Universit\'es Aix-Marseille et Sud Toulon-Var,
Luminy Case 907, 13288 Marseille, France
}
\footnotetext[3]{%
 Ecole Normale Sup\'erieure de Rabat,  
BP 5118, Rabat, Morocco}
 \setcounter{footnote}{3}

 \thispagestyle{empty}



\section{Introduction}

The critical behaviour in certain spin systems, such as the Ising model, can 
be specified in two equivalent, though conceptually quite different ways.
In the absence of an external magnetic field, decreasing the temperature
leads eventually to the onset of spontaneous symmetry breaking and hence
to the singular behaviour of derivatives of the partition function. On the
other hand, the average size of clusters of like-sign spins also diverges 
at a certain temperature, i.e., there is an onset of percolation. The 
relation between these two distinct forms of singular behaviour has been 
studied extensively over the years, and it was shown that for the Ising 
model on the lattice $\mathbbm{Z}^d$, with $d\geq 2$, implemented with a 
suitable cluster definition using temperature dependent bond weights, the
two forms lead to the same criticality: the critical temperatures $T_c$ as 
well as the 
corresponding critical exponents coincide in the two formulations 
\cite{F-K,C-K}.

In the presence of an external field $H$, the $Z_2$ symmetry of the Ising
model is explicitly broken and hence there is no more thermodynamic critical
behaviour. Geometric critical behaviour persists, however; for $T\leq T_p(H)$,
there is percolation, while for $T>T_p(H)$, the average cluster size remains
finite. In the $T-H$ plane, there thus exists a line $T_p(H)$, the so-called
Kert\'esz line, separating a percolating from a non-percolating ``phase''
\cite{Ker}. Given the mentioned correct cluster definition, it starts at 
$T_p(0)=T_c$, i.e., at the thermodynamic critical point.

\begin{figure}[h]
\centerline{\psfig{file=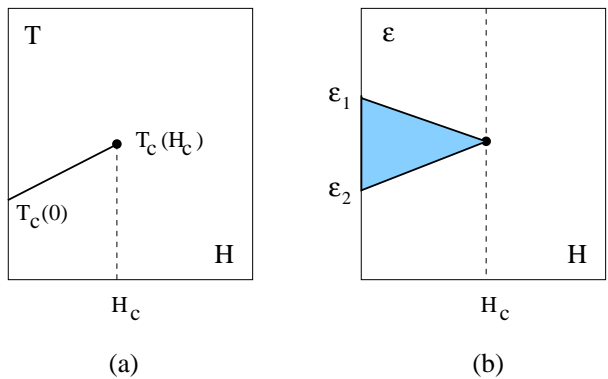,width=8cm}}
\vskip0.5cm
\centerline{\psfig{file=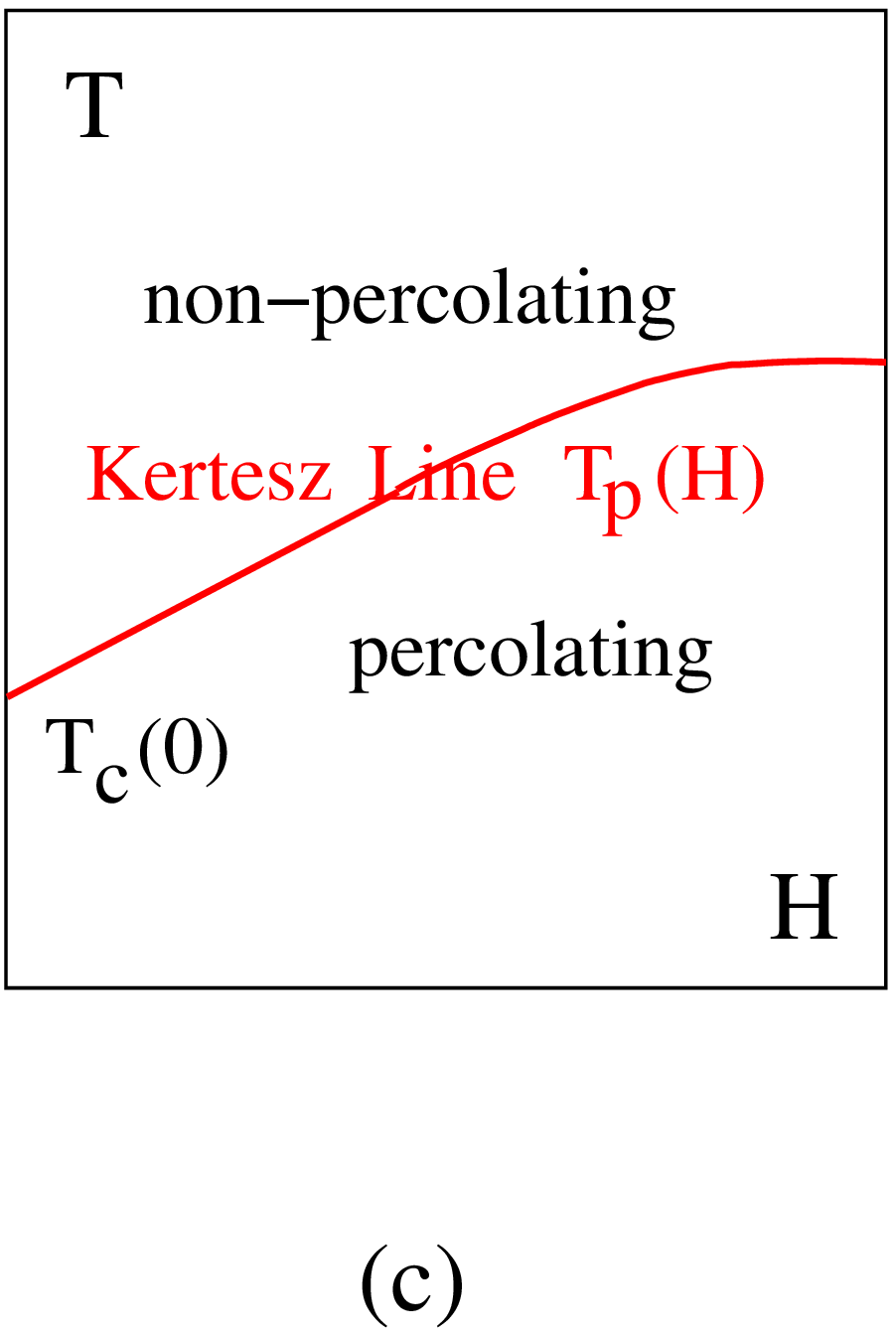,width=3.5cm}}
\medskip
\caption{Thermodynamic and geometric phase structure for a first
order transition}
\label{phase}
\end{figure}

We want to show here that the equivalence of thermodynamic and geometric
critical behaviour can be extended to the case $H\not=0$. Since in the case 
of continuous thermodynamic transitions, such as those of the Ising model,
the introduction of an external field excludes singular behaviour (for
the case $d=2$ this is shown analytically \cite{Isakov,Pfister}), our problem
makes sense only for first order transitions, for which the discontinuity
remains over a certain range of $H$, even though for $H\not=0$, the symmetry 
is broken. The ideal tool for such a study is the $q$-state Potts' model 
on a  lattice $\mathbbm{Z}^d$, with $q\geq 3$ and $d\geq 3$. In this case,
we have a thermodynamic phase diagram of the type shown in 
Fig.\ \ref{phase}a, with a line of first order transitions starting 
at $T_c(0)$ and ending at a second order point $T_c(H_c)$ \cite{K-S}; the 
transition at this endpoint is found to be in the universality 
class of the 3-d Ising model. In terms of the energy density $\epsilon(H)$ of 
the system (the energy per lattice volume), the phase diagram has the form
shown in Fig.\ \ref{phase}b; for $H=0$, the coexistence range $\epsilon_2 
\leq \epsilon(0) \leq \epsilon_1$ corresponds to the critical temperature 
$T_c(0)$. The average spin $m(\epsilon)$ as order parameter vanishes for 
$\epsilon \geq \epsilon_1
$ and becomes finite for smaller $\epsilon$.
We want to show that in the temperature range 
$T_c(0) \leq T(H) \leq T_c(H_c)$, the corresponding Kert\'esz line 
$T_p(H)$ (see Fig.\ \ref{phase}c)
coincides with that of the thermal discontinuity and that it also 
leads to the same first order/second order phase structure. Let us begin 
with a conceptual discussion of the situation. 

\vspace*{0.5cm}

\medskip

The $q$-state Potts' model in the absence of an external field provides
$q+1$ phases: the disordered phase at high temperature and $q$ degenerate
ordered low-temperature phases. Spontaneous symmetry breaking has the
system fall into one of these as the temperature is decreased. Turning on 
a small external field $H$ aligns the spins in its direction and thus 
effectively removes the $q-1$ ``orthogonal'' low-temperature phases. Hence 
now only two phases remain: the ordered low-temperature state of spins
aligned in the direction of $H$, and the disordered high-temperature phase. 
The two are for $T<T_c(H_c)$ separated by a mixed-phase coexistence regime. 
At the endpoint $T=T_c(H_c)$, there is a continuous transition from a 
system in one (symmetry broken) ordered phase to the corresponding 
(symmetric) disordered phase. The behaviour at $H=H_c$
in Fig.\ \ref{phase}b is thus just that of the Ising model, and hence
the endpoint transition is in its universality class.

In the geometric formulation for $H=0$, with decreasing temperature or 
energy density there is formation of finite clusters
of $q$ different orientations; 
the clusters here are defined using the temperature-dependent F-K bond 
weights. At $\epsilon(0)=\epsilon_1$, the $Z_q$ symmetry is spontaneously 
broken: for 
one of the $q$ directions, there now are percolating clusters, and the
percolation strength $P(\epsilon)$ becomes finite for $\epsilon<\epsilon_1$. 
However,
the disordered phase also still forms a percolating medium (for $d\geq 3$).
A further decrease of the energy density reduces the fraction of space
in disordered state, and for $\epsilon(0)\leq \epsilon_2$, there is no more 
disordered percolation. Embedded in the disordered phase are at all
times finite clusters of a spin orientation ``orthogonal'' to the one
chosen by spontaneous symmetry breaking. In our treatment, we will
therefore divide the set of clusters into three classes: disordered,
ordered in the direction of symmetry breaking, and ordered orthogonal to the
latter. While for $H=0$, any of the $q$ directions could be the 
given orientation, for $H\not=0$, the external field specifies the
alignment direction, making the $q-1$ sets of ``orthogonal'' clusters
essentially irrelevant. It is for this reason that at the endpoint
of a line of first order transitions one generally encounters the
universality class of the Ising model. Whatever the original symmetry
of the system was, at the endpoint there remains only the aligned
and the disordered ground states.

The plan of the paper is as follows. In the next section, we recall
the cluster treatment of the Potts' model and specify our method to
identify the different cluster types. This will be followed by an
analytic study valid for small external fields and by numerical
calculations for different $q$ up to asymptotic values of $H$.
Formal details of the analytic calculation are given in the appendix.

\section{The model}

We consider a finite--volume  $q$--state Potts model on the lattice 
$\mathbbm{Z}^d$ ($d \geq 2$), at inverse temperature $\beta=1/T$ and 
subject to an external ordering field $h$. It is defined by the 
Boltzmann weight
\begin{equation}
  \omega_{\tmop{Potts}} (\tmmathbf{\sigma}) 
= 
\prod_{\langle i, j
  \rangle} e^{\beta ( \delta_{\sigma_i, \sigma_j} - 1 )} \prod_i
  e^{h \delta_{\sigma_i, 1}}, 
\label{Potts}
\end{equation}
where the spins $\sigma_i$ take on the values of the set $\{ 1,\ldots, q \}$, 
and where the first product is over nearest neighbour pairs (n,n). If we 
want to study the behaviour of clusters, in the sense of F-K clusters,
we turn to the corresponding Edwards--Sokal formulation {\cite{ES}}, given 
by the Boltzmann weight
\begin{equation}
  \omega_{\tmop{ES}} (\tmmathbf{\sigma},\tmmathbf{\eta}) 
= 
\prod_{\langle i, j
  \rangle} [ e^{- \beta} \delta_{\eta_{i j}, 0} + ( 1 - e^{- \beta} )
  \delta_{\eta_{i j}, 1} \delta_{\sigma_i, \sigma_j} ] 
\prod_i e^{h \delta_{\sigma_i, 1}} ,
\label{ES}
\end{equation}
where the edge variables $\eta_{i j}$ belong to $\{ 0, 1 \}$. This 
``site-bond'' model can be thought of as follows. Given a certain spin 
configuration, one puts between two neighbouring sites $\sigma_i = \sigma_j$
an edge or bond with the probability $1-e^{-\beta}$, and no edge with the
probability $e^{-\beta}$; for  $\sigma_i \ne \sigma_j$, no bond is present.  
When the field is infinite, all $\sigma_i=1$, and we are left with a classical bond percolation problem, while for finite field, one has a random bond percolation model in the random media given by the spin configuration.

In the presence of an external field,  we find it convenient for the study 
of the Kert\'esz's line to consider a modified version of the
Edwards--Sokal formulation. We have three different types of spin
combination: (0) two adjacent spins $i,j$ are not equal, (1) two adjacent
spins $i,j$ are equal and parallel to $h$ (we denote this direction as 1),
or (2) two adjacent spins $i,j$
are equal but not parallel to $h$. Correspondingly, we ``color'' the
edge between $i$ and $j$ in three different colors $n_{i,j}$, where the 
edge variables $n_{i j}$ belong to $\{0, 1, 2 \}$. The resulting
Boltzmann weight becomes
\begin{multline}
\omega_{\tmop{CES}} (\tmmathbf{\sigma},\tmmathbf{n}) =
\prod_{\langle i,
  j \rangle}^{_{}} [ e^{- \beta} \delta_{n_{i j}, 0} + ( 1 - e^{- \beta} )
  \delta_{n_{i j}, 1}  \chi_{(\sigma_i=\sigma_j=1)}  \\
   + ( 1 - e^{- \beta} ) \delta_{n_{i j}, 2}  
\chi_{(\sigma_i=\sigma_j \ne 1)}   ] \prod_i e^{h
  \delta_{\sigma_i, 1}},  
\label{CES}
\end{multline}
where the characteristic function $\chi (\sigma_i=\sigma_j=1)$ is unity
for $\sigma_i=\sigma_j=1$ (parallel spins in the direction of $h$) and zero 
otherwise, while $\chi (\sigma_i=\sigma_j \ne 1)$ is unity for parallel 
spins not in the direction of $h$ and zero otherwise. 
The summation over the spin variables then leads to the following 
Tricolor--Edge--Representation
\begin{multline}
  \omega_{\tmop{TER}} (\tmmathbf{n}) 
= 
\prod_{\langle i, j \rangle}
   e^{- \beta \delta_{n_{i j}, 0}} 
 ( 1 - e^{- \beta} )^{(\delta_{n_{i j}, 1} + \delta_{n_{i j}, 2} )} \times \\
e^{h S_1 (\tmmathbf{n})} ( q - 1 )^{C_2 (\tmmathbf{n})} 
( q - 1 + e^h)^{| \Lambda | - S_1 (\tmmathbf{n}) - S_2 (\tmmathbf{n})}.  
\label{TER}
\end{multline}
Here, $S_1 (\tmmathbf{n})$ and $S_2 (\tmmathbf{n})$ denote the
number of sites that belong to edges of color $1$ and color $2$,
respectively, while $C_2(\tmmathbf{n})$ denotes the number of connected 
components of the set of edges of color $2$, and $| \Lambda |$ is the number 
of sites of the lattice under consideration. 

Let us mention that such kind of graphical representation has already been considered for various spin models in presence of an external field \cite{CM}.

Let  $p_{\Lambda} ( i \leftrightarrow j )$ be the probability that the site  
$i$ is connected to $j$ by a path of edges of color $1$.
As  (geometric) order parameter we will consider the following mass--gap
(inverse correlation length)
\begin{equation}
  m ( \beta, h ) 
=
- \lim_{|i - j| 
\rightarrow \infty} \frac{1}{|i - j|} \ln
  \lim_{\Lambda \uparrow \mathbbm{Z}^d}  
p_{\Lambda} ( i \leftrightarrow j )
\label{massgap}
\end{equation}
where  $i$  and $j$ belong to some line parallel to an axis of the lattice.
As (thermodynamic) order parameter, we shall consider the mean energy
$E(\beta, h)= - \frac{1}{\beta} \frac{\partial}{\partial \beta} f(\beta, h)$,
where $f(\beta , h)$ is the free energy of the model \footnote{
Note that all partitions functions, and hence the free energies, of models (\ref{Potts} )--(\ref{TER} ) coincide
}.

\section{Analytic results}
Let us first have a look at the diagram of ground state configurations
of the TER representation which are the translation invariant configurations maximizing the Boltzmann weight (\ref{TER}).

For the color $a = 0, 1, 2$, let $b_a$  be the value of the Boltzmann weight 
of the ground state configuration of color $a$ 
per unit site. One finds
$b_0 = e^{- \beta d} ( q - 1 + e^h ), b_1 
=
( 1 - e^{- \beta} )^d e^h, b_2 = ( 1 - e^{- \beta} )^d$.

Notice that $b_0 = b_1$ on the line 
\begin{equation}
\beta_0 ( h) = \ln [ 1 + ( 1 + ( q - 1 ) e^{- h} )^{1 / d} ]
\end{equation}
and that  $b_0 = b_1=b_2$ at the point $\beta_0 ( 0)$.

\begin{figure}[h]
\begin{center}
\epsfig{file=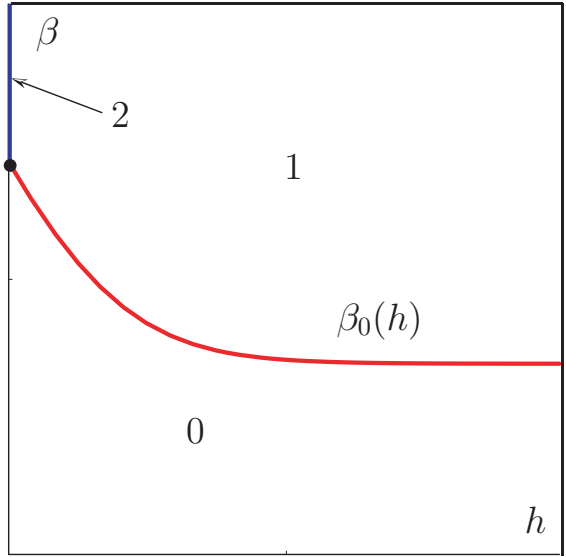}
\end{center}
\caption{Diagram of ground state configurations: 
All the ground state configurations coexist at $(0,\beta_{0}(0))$. Below $\beta_{0}(h)$, 
the $0$--state  dominates. Above $\beta_{0}(h)$, the $1$--state  
dominates; it coexists with the $0$--state on the line $\beta_{0}(h)$, and 
with the $2$--state on the line $h=0$, $\beta\geq\beta_{0}(0)$.
}
\label{gsfig}
\end{figure}

The diagram of ground state configurations, inferred from the values of the weights 
$b_0,b_1, b_2$ 
is shown in Fig.\ \ref{gsfig} (in the  $( h, \beta )$ plane). 

When  $q$ is large enough and $h$ not too large, the TER representation
(\ref{TER}) can be analyzed rigorously by a perturbative approach. 
Namely, by using the standard machinery of Pirogov--Sinai theory, we will show that the model 
undergoes a thermodynamic first order  phase transition in the sense that 
the mean energy (as well as the magnetization) is discontinuous at some  $\beta_c (h) \sim \beta_0 (h)$.
We also find for 
these values of the parameters, that the phase diagram of this model reproduces 
the diagram of ground state configurations (Fig.\ \ref{gsfig}),
see Appendix for more details.

In addition, the model exhibits a geometric 
(first order) transition, in the sense that, on the same critical line, the mass 
gap is discontinuous.

\begin{theorem}
  Assume 
$d \geq 2$,  $q$ and $h$ such that 
\begin{equation}
c_d ( 1 + ( q - 1 ) e^{- h} )^{- 1 / 2 d} < 1
\label{cond}
\end{equation} 
holds, where  $c_d$ is a given number (depending only on the dimension), 
then there exists a unique 
$\beta_c ( h ) = \beta_0 ( h ) + O (1 + ( q - 1 ) e^{- h} )^{- 1 / 2 d} )$ 
such that

\begin{enumerate}
\item 
$\Delta E(\beta_c(h),h)= E(\beta^-_c(h),h)-E(\beta^+_c(h),h)>0$
\item 
 $m ( \beta, h ) > 0$ for $\beta \leq \beta_c ( h )$ and 
$m ( \beta, h ) = 0$ for $\beta > \beta_c ( h)$.
\end{enumerate}

\end{theorem}
The proof is given in the appendix.

Let us recall that it has already been shown that  the Potts model (\ref{Potts}) undergoes, for $q$ large and 
$h$ small, a first order phase transition on a critical line \cite{BBL}, where both the mean energy and the magnetization are discontinuous.
Since, as already mentioned, the free energies of models (\ref{Potts}) and (\ref{TER}) are the same,  this critical line coincides with the one mentioned in the theorem. 

In the absence of an external magnetic field, the statements of the theorem have been  shown previously
\cite{LMMRS,LMR,KLMR}.

Condition (\ref{cond}) restricts the range of values of parameters to which
our rigorous analysis applies. Moreover, we do not expect thermodynamic
first order transitions when $h$ is sufficiently enhanced. In the next
section, we turn to numerical study on a wider range of values.

\section{Numerical simulations}

We have implemented a generalization of the Swendsen--Wang algorithm for our 
colored Edwards--Sokal model (\ref{CES}).

First, given a spin configuration, we put between any two neighbouring spins
of the same color, an edge colored  $0$ with probability (w.p.) $e^{-\beta}$, and  
w.p. $1-e^{-\beta}$, an edge colored $1$ if these spins are of 
color $1$, and colored $2$ otherwise. When two neighbouring spins disagree, 
the corresponding edge is colored $0$. 

Then, starting from an edge configuration, a spin configuration is constructed 
as follows. Isolated sites (endpoints of $0$--bonds only) are colored $1$ 
w.p.\ $e^h /(q-1+e^h)$ and colored $c\in \{2,...,q\}$ w.p.\ $1/(q-1+e^h)$. 
Non--isolated sites are colored $1$ (w.p.\ 1) if they are endpoints of
$1$--bonds and colored $c\in \{2,...,q\}$ w.p. $1/(q-1)$.

This algorithm allows us two compute both quantities associated to spins configurations and those associated to edges configurations.

\begin{figure}[h]
\begin{center}
\epsfig{file=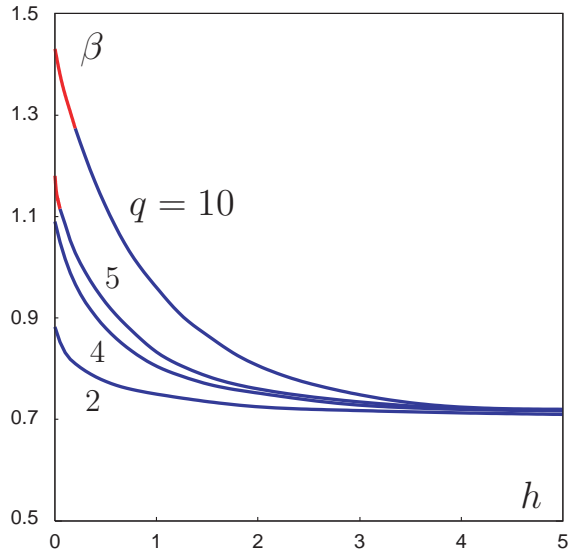}
\end{center}
\caption{$\beta_c(h)$ for several values of $q$, with ``first order'' behavior 
in red, ``second order'' in blue. The first order behavior is both thermodynamic and geometric. The second order behavior is only geometric. }
\label{fig3}
\end{figure}

The numerical results for $d=2$ are presented in Fig.\ \ref{fig3}. 
For $q\leq4$, we found a whole geometric transition line $\beta_c(h)$ for which 
$m ( \beta, h ) > 0$ when $\beta < \beta_c ( h )$, 
and $m ( \beta, h ) = 0$ when $\beta  \geq \beta_c ( h)$.
The mass gap is continuous at
$\beta_c ( h )$. For $\beta < \beta_c(h)$, the mean cluster sizes remain 
finite, while for $\beta \geq \beta_c (h)$ the size of $1$--edge clusters
diverges. 
The energy density as well as the magnetization do not show any singular behavior.

For $q\geq5$, some critical $h_c$ appears for which the geometric transition line $\beta_c(h)$ becomes first order when $h<h_c$, i.e. $m ( \beta, h ) > 0$ for $\beta \leq \beta_c ( h )$ and $m ( \beta, h ) = 0$ for $\beta > \beta_c ( h)$. In addition, on this part of the line, we find that the mean energy is discontinuous \footnote{The Swendsen-Wang algorithm allows to compute both associated order parameters (mass-gap and mean energy).}.
When $h \geq h_c$, only a geometric transition occurs and the scenario is the same as for  $q\le 4$.
Thus our numerics  show that the geometric and thermodynamic transitions coincide up to $h_c$, similarly to what we got analytically but only at (very) small field (and large $q$), see Fig.\ \ref{fig3}.

Let us mention that the numerics are in accordance with the theory for vanishing and infinite 
fields: $\beta_c(0)=\ln(1+\sqrt{q})$ and $\beta_c(\infty)=\ln 2$.

The system size in these calculations was $L=50$, $d=2$. 
The ``first order'' part of the transition lines has been 
determined via Binder cumulants \cite{BL}. The Hoshen-Kopelman algorithm
\cite{HK} was used to study cluster statistics.  For each value of $q$, 
more than $2 \times 10^5$ iterations were performed. Data have been binned 
in order to  control errors in measurements. 

\section{Concluding remarks}

For the Potts model in the presence of an external magnetic field, we have shown
that when the Kert\'esz line is first order, it coincides with the usual 
thermodynamic critical line. 
This property holds up to some critical point $(h_c, \beta_c (h_c))$,
beyond which the thermodynamic transition disappears. 
Such behavior may well appear also for a broader class of models exhibiting 
first order transition in the presence of an external field. 
We believe that the behavior at the above critical point also belongs to the 
universality class of the Ising model, as it is the case in the $3$--state 
Potts model in three dimensions \cite{K-S}.

\section{Acknowledgements}

It is a pleasure to thank J\'anos Kert\'esz for interesting comments. The BIBOS research center (Bielefeld) and the Centre de Physique Th\'eorique (CNRS Marseille) are gratefully acknowledged for warm hospitality and financial support.
The authors are indebted to an anonymous referee for constructive  remarks.

\section{Appendix}

We first introduce the partition function of the TER representation with 
boundary conditions 
$a \in \{ 0, 1, 2 \}$ 
in a box 
$\Lambda$
\footnote{The reader should not be confused by the fact that (\ref{pf}), 
called diluted partition function in PS--theory, differs from the usual 
one by an unimportant boundary term which makes the expansions (\ref{ind}) 
and (\ref{pfcm}) easier to write.}:
\begin{equation}
  Z_a ( \Lambda ) = \sum_{\tmmathbf{n}} \prod_{i \in \Lambda} \omega_i
  (\tmmathbf{n}) q^{C_2 (\tmmathbf{n}) - \delta_{a, 2}} \prod_{i \in \partial
  \Lambda} \prod_{j \sim i} \delta_{n_{i j}, a} \label{pf}
\end{equation}
where the sum is over all configurations 
$\tmmathbf{n}= \{ n_{i j} \}_{i j \cap \Lambda \emptyset}$, $\partial \Lambda$ 
is the boundary of $\Lambda$ 
(set of sites of $\Lambda$ with a n.n. in 
$\mathbbm{Z}^d \setminus \Lambda$), 
the notation $i \sim j$ 
means that $i$ and $j$ are n.n., and
\begin{equation}
  \omega_i (\tmmathbf{n}) 
= 
( 1 - e^{- \beta} )^{( \delta_{n_{i j}, 1} +
  \delta_{n_{i j}, 2} ) / 2} e^{- \beta \delta_{n_{i j}, 0} / 2} e^{h
  \chi ( i \in `` 1'' )} 
( q - 1 + e^h)^{\prod_{j \sim i} \delta_{n_{i j}, 0}} 
\label{poids}
\end{equation}
where $\chi ( i \in ``1'' )$ means that the site $i$ belongs to some 
edge of color $1$. Next, consider a configuration$\tmmathbf{n}$ 
on the envelope of $\Lambda$: $E ( \Lambda ) = \{ \langle i, j
\rangle \cap \Lambda \neq \emptyset \}$. 
A site $i \in \Lambda$ is called {\tmem{correct}} if for all 
$j \sim i$, $n_{i j}$ takes the same value, and called {\tmem{incorrect}} 
otherwise. Denote $I (\tmmathbf{n})$ the set of incorrect sites of
the configuration $\tmmathbf{n}$. A couple 
$\Gamma = \{ \tmop{Supp}
\Gamma,\tmmathbf{n}( \Gamma ) \}$ 
where the support of $\Gamma$ 
($\tmop{Supp} \Gamma$) 
is a maximal
connected subset of 
$I (\tmmathbf{n})$, and
$\tmmathbf{n}( \Gamma )$ 
the restriction of 
$\tmmathbf{n}$ 
to the envelope of
$\Lambda$ 
is called {\tmem{contour}} of the configuration $\tmmathbf{n}$ (here, a set of
sites is called connected if the graph that joins all the sites of this set at
distance 
$d ( i, j ) = \max_{k = 1, \ldots ,d} |i_k - j_k | \leq 1$ 
is connected).
A couple 
$\Gamma = \{ \tmop{Supp} \Gamma,\tmmathbf{n}( \Gamma )
\}$ where $\tmop{Supp} \Gamma$ 
is a connected set of sites is called contour
if there exists a configuration 
$\tmmathbf{n}$
 such that 
$\Gamma$ 
is a contour
of 
$\tmmathbf{n}$. 
For a contour 
$\Gamma$, 
let 
$\tmmathbf{n}_{\Gamma}$ 
denote
the configuration having $\Gamma$ as unique contour, 
$\tmop{Ext} \Gamma$
denotes the unique infinite component of 
$\mathbbm{Z}^d \setminus ( \tmop{Supp}
\Gamma )$, 
$\tmop{Int} \Gamma =\mathbbm{Z}^d \setminus ( \tmop{Ext} \Gamma
\cup \tmop{Supp} \Gamma )$, 
and 
$\tmop{Int}_m \Gamma$ 
denote the set of sites
of 
$\tmop{Int} \Gamma$ 
corresponding to the color 
$m \in \{ 0, 1, 2 \}$ 
for the
configuration 
$\tmmathbf{n}_{\Gamma}$. 
Two contours $\Gamma_1$ and $\Gamma_2$
are said to be compatible if their union is not connected and 
are called external contours if
furthermore 
$\tmop{Int} \Gamma_1 \subset \tmop{Ext}_{\Lambda} \Gamma_2$ 
and
$\tmop{Int} \Gamma_2 \subset \tmop{Ext}_{\Lambda} \Gamma_1$. 
For a family
$\theta = \{ \Gamma_1, \ldots, \Gamma_n \}_{\tmop{ext}}$ 
of external contours,
let $\tmop{Ext}_{\Lambda} \theta$  
denote the intersection 
$\Lambda \cap_{k =1}^n \tmop{Ext}_{\Lambda} \Gamma_k$. 
With these definitions and notations, one
gets the following expansion of the partition functions over families of
external contours, 
\begin{equation}
  \label{ind} Z_a ( \Lambda ) 
= \sum_{\theta 
= 
\{ \Gamma_1, \ldots, \Gamma_n
  \}_{\tmop{ext}}} b_a^{| \tmop{Ext}_{\Lambda} \theta |} \prod_{k = 1}^n \rho
  ( \Gamma_k ) \prod_{\text{$m = 0, 1, 2$}} Z_m ( \tmop{Int}_{_m} \Gamma_k ),
\end{equation}
where
$
  \label{poidscont} \rho ( \Gamma) 
=
 \prod_{i \in \tmop{supp} \Gamma}
  \omega_i (\tmmathbf{n}_{\Gamma} ) q^{C (\tmmathbf{n}_{\Gamma} ) - \delta_{a,
  2}}
$.
From
(\ref{ind}), we get
\begin{equation}
    Z_a ( \Lambda ) 
=
b_a^{| \Lambda |}
 \sum_{\{ \Gamma_1, \ldots,
    \Gamma_n \}_{\tmop{comp}}} \prod_{k = 1}^n z_a ( \Gamma_k ),
\label{pfcm}
\end{equation}
where the sum is now over families of compatible contours and the activities
$z_a ( \Gamma)$ 
of contours are given by
$z_a ( \Gamma) 
= 
\rho ( \Gamma ) b_a^{- | \tmop{supp} \Gamma |}
   \prod_{m \neq a} \frac{Z_m ( \tmop{Int}_{_m} \Gamma )}{Z_a (
   \tmop{Int}_{_m} \Gamma)}
$.

It is  easy to prove the following Peierls' estimate,
\begin{equation}
  \rho ( \Gamma ) ( \max_{a = 0, 1, 2} b_a )^{- | \tmop{Supp} \Gamma |}
  \leq 
e^{- \tau | \tmop{Supp} \Gamma |}, 
\label{Peierls}
\end{equation}
where  
$e^{- \tau} = ( 1 + ( q - 1 ) e^{- h} )^{- 1 / 2 d}$.
Indeed, first notice that an incorrect site $i$ is either of color $1$ or of
color $2$. 
In the first case one has 
$\text{$\sum_{j \sim i} ( \delta_{n_{i
j}, 0} + \delta_{n_{i j}, 1} ) = 2 d$}$, 
so that 
$\omega_i
(\tmmathbf{n}_{\Gamma} ) / b_1 = ( e^{\beta} - 1 )^{- ( \sum_{j \sim i}
\delta_{n_{i j}, 0} ) / 2}$, 
implying 
$$\omega_i (\tmmathbf{n}_{\Gamma} ) /
\max_{a = 0, 1, 2} b_a \leq ( 1 + ( q - 1 ) e^{- h} )^{- ( \sum_{j \sim i}
\delta_{n_{i j}, 0} ) / 2 d}.$$ 
Thus since 
$1 \leq \sum_{j \sim i} \delta_{n_{i j}, 0} \leq 2 d - 1$,
each incorrect site of color $1$ gives at
most a contribution 
$e^{- \tau}$ 
to the L.H.S. of (\ref{Peierls}). 
In the
second case, one has 
$\sum_{j \sim i} ( \delta_{n_{i j}, 0} + \delta_{n_{i j}, 2} ) = 2 d$, 
so that 
$w_i (\tmmathbf{n}_{\Gamma} ) / b_2 =
( e^{\beta} - 1 )^{- ( \sum_{j \sim i} \delta_{n_{i j}, 0} ) / 2},$ 
implying
$$\omega_i (\tmmathbf{n}_{\Gamma} ) / \max_{a = 0, 1, 2} b_a \leq ( q - 1 + e^h
)^{- ( \sum_{j \sim i} \delta_{n_{i j}, 0} ) / 2 d}.$$
We then use again that
$1 \leq \sum_{j \sim i} \delta_{n_{i j}, 0} \leq 2 d - 1$
and that
 $C_2 (\tmmathbf{n}_{\Gamma} ) \leq \sum_{i \in \tmop{Supp} \Gamma}
\chi ( 1 \leq \delta_{n_{i j}, 2} ) / 2^{\sum_{j\sim i}\delta_{n_{i j}, 2}}$ 
(see {\cite{KLMR}}) to obtain that each incorrect site of color 
$2$ 
gives at most
a contribution 
$( e^h + q - 1 )^{- 1 / 2 + 1 / 2 d} \leq e^{- \tau}$ 
to the
L.H.S. of (\ref{Peierls}).

When the assumptions of the theorem are satisfied, the Peierls' 
estimate (\ref{Peierls}) provides a good control of the
system by using Pirogov--Sinai theory
{\cite{S}}. We introduce the truncated activity
\[ z_a' ( \Gamma ) = \left\{\begin{array}{l}
     z_a ( \Gamma ) \tmop{if} z_a ( \Gamma ) \leq ( c_0 e^{- \tau} )^{|
     \tmop{Supp} \Gamma |}\\
     ( c_0 e^{- \tau} )^{| \tmop{Supp} \Gamma |} \tmop{otherwise,}
   \end{array}\right. \]
where $c_0$ is a numerical constant, and we call a contour stable if 
$z_a ( \Gamma
) = z_a' ( \Gamma )$. 
Let $Z_a' ( \Lambda )$ 
be the partition function
obtained from (\ref{pfcm}) by leaving out unstable contours, i.e., by taking
the activities 
$z_a' ( \Gamma )$ 
in (\ref{pfcm}), and let us  introduce the metastable
free energies
$f^{\tmop{met}}_a ( \beta, h ) = - \lim_{\Lambda \uparrow
\mathbbm{Z}^d}(1/|\Lambda |) \ln Z_a' ( \Lambda )$. 
The leading term of these metastable free energies  equals
$- \ln b_a$. The corrections can be expressed by free energies of 
contour models which can be controlled by
convergent cluster expansions. As a standard result of Pirogov-Sinai
theory, one gets that the phase diagram of the system is a small perturbation
of the diagram of ground state configurations. 
Namely, there exits a unique
point 
$\beta_c ( 0 )$ 
given by the solution of 
$f_0^{\tmop{met}} ( \beta, h )
= f_1^{\tmop{met}} ( \beta, h ) = f_2^{\tmop{met}} ( \beta, h )$ 
for which all
contours are stable and such that 
$Z_a ( \Lambda ) = Z_a' ( \Lambda )$ 
for 
$a
= 0, 1, 2$. 
There exists a line 
$\beta_c ( h )$ 
given by the solution of
$f_0^{\tmop{met}} ( \beta, h ) = f_1^{\tmop{met}} ( \beta, h )$  
when 
$h > 0$
and such that,  
$Z_a ( \Lambda ) = Z_a' ( \Lambda )$ for $a = 0, 1$. 
For
$\beta < \beta_c ( h )$ 
one has $Z_0 ( \Lambda ) = Z_0' ( \Lambda )$, 
and for
$\beta > \beta_c ( h )$ 
one has $Z_1 ( \Lambda ) = Z_1' ( \Lambda )$. 
For 
$h = 0$ and $\beta \geq \beta_c ( 0 )$, 
one has in addition $Z_2 ( \Lambda ) = Z_2'
( \Lambda )$.

For the color $a=0,1,2$, denote by 
$\langle \cdot \rangle^a (\beta , h)$
the expectation value under the $a$--boundary condition.
As a consequence of the above expansions and analysis, we obtain by standard Peierls' estimates that for $h\geq 0$
\begin{eqnarray}
\langle \delta_{n_{i j},1} \rangle^1 (\beta , h)
&\geq &
1-O(e^{- \tau}) \quad \text{for} \quad \beta \geq \beta_c (h)
\\
\langle \delta_{n_{i j},0} \rangle^0 (\beta , h)
&\geq &
1-O(e^{- \tau}) \quad \text{for} \quad \beta \leq \beta_c (h)
\end{eqnarray}
while in addition we also get for $h=0$:
\begin{equation}
\langle \delta_{n_{i j},2} \rangle^2 (\beta , 0)
\geq 
1-O(e^{- \tau}) \quad \text{for} \quad \beta \geq \beta_c (0)
\end{equation}
By definition of the mean energy, one  has that  $\Delta E=E(\beta^-,h)-E(\beta^+,h)$ is proportional to the difference 
$\langle \delta_{n_{i j},0} \rangle^0 (\beta^- , h)-\langle \delta_{n_{i j},0} \rangle^1 (\beta , h)$, and 
the first statement of the theorem  follows immediately from these properties.

To prove the second statement,
we remark that if one imposes that the site $i$ is
connected to $j$ by a path made up of edges of color $1$, 
then under the boundary
condition $0$, 
there exists necessarily an external contour that encloses both
the sites $i$ and $j$. 
As a consequence of the above analysis the probability
of external contours $\Gamma$ decays like 
$(c_0 e)^{- \tau | \tmop{Supp} \Gamma |}$
when the $0$--contours are stable, i.e. when 
$Z_0 ( \Lambda ) = Z_0' ( \Lambda) $. 
One thus  gets 
$p_{\Lambda} ( i \leftrightarrow j ) \leq ( \tmop{Cte} e^{- \tau} )^{|i - j|}$ 
when 
$\beta \leq \beta_c ( h )$ 
from which  the first statement of the
theorem follows. 
On the other hand under the boundary condition $1$, the probability
that the site $i$ is not connected to $j$ can be bounded from above by a small
number $O ( e^{- \tau} )$ when 
$Z_1 ( \Lambda ) = Z_1' ( \Lambda ) $. 
This follows  also from a Peierls type arguments and 
implies that the probability that
the site $i$ is connected to $j$ under the boundary condition $1$ is greater
than 
$1 - 0 ( e^{- \tau} )$ 
for $\beta \geq \beta_c ( h )$. 
It gives also that the
probability 
$p_{\Lambda} ( i \leftrightarrow j )$ 
for the site $i$ to be connected with $j$
under the boundary condition $0$ is also greater than 
$1 - 0 ( e^{- \tau} )$
for $\beta > \beta_c ( h )$, implying the second statement.

\end{document}